\documentclass[conference]{ieeetran}  
\IEEEoverridecommandlockouts
\usepackage{cite}
\usepackage{amsmath,amssymb,amsfonts}
\usepackage{algorithmic}
\usepackage[ruled,vlined,linesnumbered]{algorithm2e}
\usepackage{blkarray}
\usepackage{bbm}
\usepackage{bm}
\usepackage[makeroom]{cancel}
\usepackage{cases}
\usepackage[usenames,dvipsnames]{color}
\usepackage{dsfont}
\usepackage{epsfig}
\usepackage[T1]{fontenc}
\usepackage{graphicx}
\usepackage{flushend}
\usepackage{textcomp}
\usepackage{xcolor}
\usepackage[UKenglish,english]{babel}
\usepackage{amsmath,amssymb,paralist,subfigure,graphicx,stmaryrd,mathrsfs,url,algorithm2e,soul}
\usepackage{cuted,mathtools,lipsum,cite}
\usepackage{float}
\usepackage{flushend}
\usepackage{graphicx}
\usepackage{hyperref} 
\usepackage{mathtools}
\usepackage{mathrsfs}
\usepackage{setspace}
\usepackage{soul} 
\usepackage{subfigure}
\usepackage{stfloats}
\usepackage{tikz}
\usepackage{verbatim}
\usepackage{xspace}

\def\mb{\mathbf}

\def\mc{\mathcal}

\begin{document}
\title{\LARGE \bf Distributed Finite-Sum Constrained Optimization subject to Nonlinearity on the Node Dynamics    } 

\author{\IEEEauthorblockN{Mohammadreza Doostmohammadian$^{1,2}$, Maria Vrakopoulou$^{3}$,
 Alireza Aghasi$^{4}$, Themistoklis Charalambous$^{1,5}$}
\IEEEauthorblockA{$^{1}$\textit{School of Electrical Engineering at Aalto University, Espoo, Finland, 
\texttt{name.surname@aalto.fi}}
\\ $^{2}$\textit{Mechatronics Department, Faculty of Mechanical Engineering at Semnan University,  Iran, \texttt{doost@semnan.ac.ir}}
\\ $^{3}$\textit{University of Melbourne, Australia, email: \texttt{maria.vrakopoulou@unimelb.edu.au}}
\\ $^{4}$\textit{Robinson College of Business, Georgia State University, USA, \texttt{aaghasi@gsu.edu}}
\\ $^{5}$\textit{Department of Electrical and Computer Engineering, University of Cyprus, Cyprus, \texttt{surname.name@ucy.ac.cy}}
} 
\thanks{This work is supported in part by
the European Commission through the H2020 Project Finest Twins under
Agreement 856602.}
}

\maketitle
\thispagestyle{empty}
\begin{abstract}
	Motivated by recent development in networking and parallel data-processing, we consider a distributed and localized \textit{finite-sum} (or fixed-sum) allocation technique to solve resource-constrained convex optimization problems over multi-agent networks (MANs). Such networks include (smart) agents representing an intelligent entity capable of communication, processing, and decision-making. In particular, we consider problems subject to practical \textit{nonlinear constraints on the dynamics} of the agents in terms of their communications and actuation capabilities (referred to as the node dynamics), e.g., networks of mobile robots subject to actuator saturation and quantized communication. The considered distributed sum-preserving optimization solution further enables adding purposeful nonlinear constraints, for example, sign-based nonlinearities, to reach convergence in predefined-time or robust to impulsive noise and disturbances in faulty environments. Moreover, convergence can be achieved under minimal network connectivity requirements among the agents; thus, the solution is applicable over dynamic networks where the channels come and go due to the agent's mobility and limited range. This paper discusses how various nonlinearity constraints on the optimization problem (e.g., collaborative allocation of resources) can be addressed for different applications via a distributed setup (over a network).           
\end{abstract}

\section{Introduction} \label{sec_intro}
\IEEEPARstart{D}{istributed} and localized solutions are widespread in signal-processing, machine-learning, and control literature \cite{csl2021}, and are getting more attention due to recent advancements in wireless communication devices, green networking \cite{7110517},  Internet-of-Things (IoT), cloud computing, and high-performance central processing units (CPUs). These techniques further found their way into vehicular technology, and intelligent transportation systems, e.g., Internet-of-Connected-Vehicles (IoCV) \cite{8960642,bute2021collaborative} and platoons of autonomous cars \cite{Abolfazli2021Platoon}. The concepts of computing-workload management over a network of CPUs \cite{kalyvianaki2009self,charalambous2010min,rikos2021optimal}, optimal scheduling of reserving batteries \cite{vrakopoulou2017chance}, optimal electricity generation for greener smart grid \cite{chen2016distributed,cherukuri2015tcns,chen2018fixed,molzahn2017survey,yang2013consensus}, optimal scheduling of plug-in EV charging \cite{falsone2020tracking},   
and proportional task allocation  \cite{MiadCons,MSC09,moarref2008facility,ando1999distributed} are relevant to this research area. Another example is to address the best performance of the power-constrained solutions for localized coverage and deployment control of mobile facilities and service-providing units \cite{MSC09}. The other notions being addressed are practical nonlinearities on the agents' models. For example, real actuators in mobile robots are subject to actuator saturation while considering ideal models may cause accumulated tracking error (integral windup) and poor control performance, and, in turn, more energy consumption and undesired non-optimal system output, in general. In many practical situations, moreover, it is of interest \textit{to add intended nonlinearities}, for example, to reach convergence in the fixed (or finite) amount of time \cite{garg2020fixed}, or to make the control solution robust to impulsive noise and resilient to disturbances. This paper addresses the optimality of the solutions subject to such nonlinear constraints on the agents' dynamics over a distributed network setup.  

The existing distributed optimization solutions are mainly linear (on the gradient) \cite{gharesifard2013distributed,doan2017convergence,doan2017scl} with no possibility to tolerate certain nonlinearities and to address the above-mentioned practical situations. Some existing models addressing specific nonlinearities are listed here. Quantized consensus \cite{kashyap2007quantized} with applications to finite-time CPU scheduling \cite{rikos2021optimal} (based on the results of \cite{kalyvianaki2009self,charalambous2010min}) is considered, where the cost function needs to be quadratic. Sign-based solutions \cite{MSC09,taes2020finite} are proposed to address saturation constraints and advance the linear models in coordination of intelligent mobile robots for optimal coverage control \cite{cortes2004coverage} and linear machine learning optimization methods \cite{Xin2020khan}. \textit{Accelerated} (but linear) machine learning \cite{hendrikx2019accelerated} and optimization methods (e.g., heavy-ball \cite{xin2019distributed}) along with specific nonlinear dynamics \cite{mrd_2020fast,9609591} are developed recently. In the context of synchronization, consensus, and optimal agreements, similar nonlinear constraints in real-world applications are given, e.g., in \cite{wei2018nonlinear}. For a summary of different existing constraints on the cost functions and the associated solutions, see \cite[Table I]{mrd_2020fast}. The existing literature mainly focuses on particular applications customized with a specific structure and, in general, not applicable in other problems with various nonlinear models. In other words, a general framework to address all such nonlinearities (e.g., a composition nonlinearity) is missing. Such a general solution is more practical for applications including one (or more) composite nonlinearity, e.g., reaching faster or disturbance-resilient convergence while addressing quantization or clipping effects, ranging from robotic science to machine learning.

In this paper, we consider distributed optimization solutions addressing nonlinear constraints on the model of agents based on \textit{only local information exchange}. This better suits spatially distributed MAN and networked control systems \cite{Yang2010}, e.g., mobile robotic networks and geographically distributed sensor networks. The proposed method applies to general nonlinearities, including (but not limited to) saturation and quantization actuation and communication models. We further discuss solutions other than sector-based nonlinearities, e.g., both uniformly quantized and logarithmically quantized solutions are addressed here. Another example is fast-convergent (in finite/fixed-time) by \textit{adding} non-Lipschitz nonlinearities or sign-based dynamics to suppress disturbances in designed actuation and available communication channels. The proposed solution covers constrained distributed optimization and resource allocation problems ranging from machine learning to smart grid (e.g., automatic generation control). We make the slightest assumption on the MAN connectivity, i.e., uniform connectivity over time. In other words, we ensure convergence despite losing network connectivity as far as the combination of switching network over some finite non-overlapping time-intervals is \textit{uniformly-connected}. This is irrespective of the nonlinearity-type involved with the agents' dynamics and, thus, opens up many possible application venues.  
We omit the mathematical proofs here and refer interested readers to \cite{mrd_2020fast,mrd20211st} for details. We consider Laplacian-gradient-based solutions, mainly used for convex optimization problems, to address the nonlinear models on agents or their communications over a distributed setup. For general non-convex optimization problems, as in 
\cite{MSC09,MiadCons}, such solutions might be sub-optimal, and approximation algorithms are practically applied in such cases (some of these problems are even NP-hard by nature), while they can be applied to some other specific non-convex cases \cite{xin2021improved}. Applications to CPU scheduling over computing server networks with quantized information transmission and economic dispatch for green smart grid networks with improved convergence rate are provided as some illustrative examples. 

\section{The Optimization Framework}
We consider general constrained convex optimization problems in the form,
\begin{align} \label{eq_dra}
\min_\mb{x}
~~ & F(\mb{x}) = \sum_{i=1}^{n} f_i(\mb{x}_i)~~
\text{s.t.} ~ \sum_{i=1}^{n} \mb{x}_i = K
\end{align}
which may represent more general problems obtained by scaling and change of variables. Without loss of generality, assume $\mb{x}_i,K \in \mathbb{R}$ (where it can be easily extended to $\mb{x}_i,K \in \mathbb{R}^d$). In problem \eqref{eq_dra}, the overall cost function needs to be minimized (or dual problem of maximizing utilities) and $f_i(\mb{x}_i)$ represent the local costs as a function of local resources at each agent (node) $i$. These costs are generally assumed (or could be approximated by) strongly or strictly convex and smooth functions (even some special non-convex cases can be addressed \cite{xin2021improved}). 
The single constraint $\sum_{i=1}^{n} \mb{x}_i = K$ is referred to as the \textit{feasibility} condition, and implies constant amount of total resources to be allocated among a group of agents. For example, this may represent the entire convex area covered by a robotic network or the overall load demand and the amount of power allocated for generator coordination over power grids. For different applications, the problem could be subject to some additional constraints\footnote{Recall that having multiple constraints in the form $A\mb{x}=b$, the problem \eqref{eq_dra} is not necessarily solvable locally. In this case, one can find the unique intersection of constraint faucets (if any) and reduce the constraints to a single lower-order one by presenting some states say $\mb{x}_a$ as a linear combination of other complementary states $\mb{x}_{-a}$, say $\mb{x}_{j_1},\mb{x}_{j_2},\dots,\mb{x}_{j_m}$.  However, in this form, the cost function $f_a(\mb{x}_a)$ in \eqref{eq_dra} cannot be generally decoupled as a summation of separate $\overline{f}_{j_1}(\mb{x}_{j_1}),\overline{f}_{j_2}(\mb{x}_{j_2}),\dots,\overline{f}_{j_m}(\mb{x}_{j_m})$, and, although the problem might be solvable via centralized solutions, it cannot be necessarily solved in a localized and distributed setup and is irrelevant in the sum-preserving distributed optimization literature. In \textit{some} cases, the problem formulation is decoupled in terms of costs while linearly coupled by a \textit{reduced-order} single constraint, and thus, is solvable with existing distributed solutions. Therefore, as the title says in this paper, we skip the multiple-constraint formulation (which needs further elaboration) and only focus on single-constraint finite-sum problems as in many literature. 
}. For example, the so-called \textit{box constraints} in the form,
\begin{align} \label{eq_box}
    m_i \leq \mb{x}_i \leq M_i
\end{align}
These constraints can be addressed by adding penalty functions \cite{bertsekas2003convex} to the cost function in \eqref{eq_dra} in the form,
\begin{align}
    f_i^\epsilon (\mb{x}_i) = f_i(\mb{x}_i) + \epsilon [\mb{x}_i - M_i]^+ + \epsilon [m_i - \mb{x}_i]^+
\end{align}
with $[u]^+ = \max \{u,0\}$. It can be proved that the solution of this penalized case can become arbitrary close to the exact optimizer by choosing $\epsilon$ sufficiently small. This non-smooth function can be substituted by the following smooth equivalents \cite{nesterov1998introductory,csl2021},
\begin{align}
[u]_\mu^+ &= \frac{1}{\mu} \log (1+\exp (\mu u)) \\
[u]^{+2} &= ([u]^{+})^2
\end{align}
The initialization under different $M_i,m_i$ values for different agents can be addressed via the algorithm in \cite{cherukuri2015tcns} or in some cases by (linearly) scaling or normalizing the variables via, 
\begin{align}
   \mb{z}_i = \frac{\mb{x}_i-m_i}{M_i-m_i}  
\end{align}
and scaling back the optimization variables $\mb{z}_i$ to $\mb{x}_i$ after termination and convergence. In this case, the constant $K$ needs to be adjusted accordingly, beforehand.
The feasibility constraint could be originally resulted from a \textit{weighted} sum in the form $\sum_{i=1}^{n} \mb{x}_i = \sum_{i=1}^{n} a_i\mb{y}_i = K$,
where the change of variables in the form $a_i\mb{y}_i = \mb{x}_i$ is applied. In this case, we avoid solution singularity for small $a_i$ as compared to the other coefficients $a_{-i}$, since we consider $a_i\mb{y}_i = \mb{x}_i$ instead of directly solving for $\mb{y}_i$. In general, it is typically assumed that $\frac{K}{a_i}$ and $\frac{a_j}{a_i}$ are bounded for all $i,j$. This is because following the Karush–Kuhn–Tucker (KKT) condition, for the optimizer $\mb{y}^*$, we have $\frac{\partial_y f_i(\mb{y}_i^*)}{a_i} =  \frac{\partial_y f_j(\mb{y}_j^*)}{a_j}$. This implies that $\frac{\partial_y f_j(\mb{y}_j^*)}{\partial_y f_i(\mb{y}_i^*)}$ remains bounded for all $i,j$. 

Using the KKT condition,
the unique optimal solution ${\mb{x}^*}$ for \eqref{eq_dra} is in the form
$ \nabla F({\mb{x}^*}) = {\psi}^*\mb{1}_n$,
where $\nabla F({\mb{x}^*}) = (\partial_x f_1(\mb{x}_1^*);\dots; \partial_x f_n(\mb{x}_n^*))$ denotes the gradient of $F$ at $\mb{x}^*$, and $\mb{1}_n$ is the column vector of $1$'s. 
In general, for every $K$ there is a unique $\psi^*$ and a unique optimizer $\mb{x}^*$ satisfying $ \nabla F({\mb{x}^*}) = {\psi}^*\mb{1}_n$. 
In distributed setup, this unique optimizer $\mb{x}^*$ needs to be \textit{invariant} under the solution dynamics. 
The existing solution dynamics in the literature are mainly first order in both discrete-time (DT) and continuous-time (CT) format where $\dot{\mb{x}}_i$ (or $\mb{x}_i(k+1)$) is a function of $\mb{x}_i, \mb{x}_j,\partial_x f_i,  \partial_x f_j,W$ at time $t$ (or time-step $k$) with $j\in \mc{N}_i$, 
$\partial_x f_i$ representing the local gradient, $W$ representing the (possibly time-varying) adjacency matrix of the (dynamic) MAN, and the set of in-neighbors $\mc{N}_i = \{j|j\rightarrow i\}$ (or $\mc{N}_i = \{j|W_{ji} \neq 0\}$). Each agent uses the information (local gradient and state) of its own and neighbors and its direct neighbors.
For example, linear $1$st-order solutions for \eqref{eq_dra} are in the form \cite{gharesifard2013distributed,cherukuri2015tcns},
\begin{align} \label{eq_ct_lin}
    & \dot{\mb{x}}_i = \eta \sum_{j\in \mc{N}_i}W_{ji}(t)(\partial_x f_j(t)-\partial_x f_i(t))=-L \nabla F({\mb{x}})  \\ \label{eq_dt_lin}
    & \mb{x}_i(k+1) = \mb{x}_i(k) + \overline{\eta} \sum_{j\in \mc{N}_i}W_{ji}(k)(\partial_x f_j(k)-\partial_x f_i(k)),
\end{align}
subject to some possible additional stochastic condition on $W$.  This is, further, addressed based on $L$ as the Laplacian matrix (hence, the name Laplacian-gradient method \cite{cherukuri2015tcns}).

In this work, we address some  additional constraints  on the dynamics of agents $\dot{\mb{x}}_i$ (i.e., the constraint on the actuators) and/or their communications over MAN as discussed below.
\begin{itemize}
    \item Nonlinear communication constraints: The communicated variable ($\mb{x}_j$ or $\partial_x f_j$) of agent $j\in \mc{N}_i$ goes through a nonlinear channel and agent $i$ receives $h_c(\mb{x}_j)$ or $h_c(\partial_x f_j)$ after a nonlinear mapping $h_c(\cdot)$ is applied.
    
    \item Nonlinear actuation (or node-based) constraints: The input $u_i$ in  $\dot{\mb{x}}_i = u_i$ (or $\mb{x}_i(k+1)- \mb{x}_i(k)= u_i(k)$)  is a function of $\mb{x}_j-\mb{x}_i$ or $\partial_x f_j- \partial_x f_i$ (e.g., see the right-hand-side (RHS) of \eqref{eq_ct_lin}-\eqref{eq_dt_lin}). In case of nonlinear actuation, the actuator output (to be applied as input $u_i$) is in the form $h_a(\mb{x}_j-\mb{x}_i)$ or $h_a(\partial_x f_j-\partial_x f_i)$ as compared to the originally designed input (e.g., in \eqref{eq_ct_lin}-\eqref{eq_dt_lin}). This implies a nonlinear actuation mapping defined by $h_a(\cdot)$.    
\end{itemize}
In the rest of the paper, we use the same notation $h(\cdot)$ for both nonlinearities $h_a(\cdot), h_c(\cdot)$. Possible nonlinear mappings $h(\cdot)$ in practical applications include, for example, saturation (or clipping), denoted by $h(\mb{z}) = \mbox{sat}(\mb{z})$, and quantization in the uniform form $h(\mb{z})=q_{u}(\mb{z}) = \delta \left[ \frac{\mb{z}}{\delta}\right]$ (with  $\left[ \cdot\right]$ as rounding to the nearest integer) and logarithmic form $h(\mb{z})=q_{l}(\mb{z}) = \mbox{sgn}(\mb{z}) \exp(q_{u}(\log(|\mb{z}|)))$ (with  $\mbox{sgn}(\mb{z})=\frac{\mb{z}}{|\mb{z}|}$). In many applications it is desired to reach convergence in fixed (pre-defined) time. Then, one can consider non-Lipschitz nonlinearities in the form $\mbox{sgn}^{\mu_1}(\mb{z})+ \mbox{sgn}^{\mu_2}(\mb{z})$ with $\mbox{sgn}^\mu(\mb{z})=\mb{z}|\mb{z}|^{\mu-1}$. Putting $\mu = 0$ one can get sign-based dynamics which are known to be robust against impulsive noise and disturbances, see examples of sign-based dynamics for consensus and synchronization in \cite{9030120} and references therein. 

In this paper, we study general dynamics with nonlinear actuation $h_a(\cdot)$ and/or communication $h_c(\cdot)$ involved (in terms of nonlinearity concerning the gradient) that may emerge in different practical applications or designed for specific purposes. 
We discuss the exact convergence to the optimizer for both \textit{sector-based} and \textit{non-sector-based} model nonlinearities and determine the $\varepsilon$-neighborhood of the optimizer in problems for which the exact convergence cannot be guaranteed.
Besides the nonlinearities mentioned above, our results can address their composition.

\section{The General Distributed Solution} \label{sec_solution}
We extend the linear DT solution \eqref{eq_dt_lin} to account for both nonlinearity in communications and actuation. The general nonlinear solutions are in the form,

\small \begin{align} \label{eq_dt_nonlin1}
    \mb{x}_i(k+1) &= \mb{x}_i(k) + \overline{\eta} \sum_{j\in \mc{N}_i}W_{ji}(k)h_a(\partial_x f_j(k)-\partial_x f_i(k)), \\ \label{eq_dt_nonlin2}
    \mb{x}_i(k+1) &= \mb{x}_i(k) + \overline{\eta} \sum_{j\in \mc{N}_i}W_{ji}(k)(h_c(\partial_x f_j(k))-h_c(\partial_x f_i(k))),
\end{align} \normalsize
Similar nonlinear solutions can be stated for CT protocol \eqref{eq_ct_lin}. 
\begin{align} \label{eq_ct_nonlin1}
    \dot{\mb{x}}_i &=  \eta\sum_{j\in \mc{N}_i}W_{ji}(t)h_a(\partial_x f_j(t)-\partial_x f_i(t)), \\ \label{eq_ct_nonlin2}
     \dot{\mb{x}}_i &=  \eta \sum_{j\in \mc{N}_i}W_{ji}(t)(h_c(\partial_x f_j(t))-h_c(\partial_x f_i(t))),
\end{align} \normalsize
Some relevant examples are given in \cite{mrd_2020fast,mrd20211st}.
The RHS of \eqref{eq_dt_nonlin1}-\eqref{eq_dt_nonlin2} could be discontinuous due to dynamic network topology. In that sense, notions of discontinuous systems and generalized gradients can be applied for Lyapunov convergence analysis.  

First, we discuss the feasibility condition for the given protocols, i.e., in practical applications (e.g., generator coordination for economic dispatch), the solution needs to remain sum-invariant. This is necessary for convergence to the exact optimizer of \eqref{eq_dra} and for non-feasible protocol there is no guarantee that the solution converges to the same $\mb{x}^*$ satisfying $ \nabla F({\mb{x}^*}) = {\psi}^*\mb{1}_n$. 
In mathematics form, the protocols \eqref{eq_dt_nonlin1}-\eqref{eq_dt_nonlin2} need to satisfy $\sum_{i=1}^n \mb{x}_i(k+1) = \sum_{i=1}^n \mb{x}_i(k)$. It can be shown that \cite{mrd_2020fast,mrd20211st} for odd nonlinearities (i.e., $h(-z) = -h(z)$) and \textit{bidirectional links} over MAN the feasibility condition holds. This implies that for all symmetric nonlinear mapping $h(\cdot)$ the feasibility condition holds and, thus, the optimizer $\mb{x}^*$ is the invariant solution of the proposed dynamics. All the above mentioned nonlinearities, e.g., saturation and quantization, are symmetric for both positive and negative variables. Recall that in many applications, the agents have similar communication devices in terms of range. This implies that the bidirectional links over homogeneous MAN is well-justified. Moreover, for protocol \eqref{eq_dt_nonlin2} the solution can be further extended to \textit{strongly-connected with directed communications} over MAN under certain design of the weighting (consensus) matrix $W$, referred to as \textit{doubly-stochastic} condition or \textit{balanced network} design \cite{csl2021}.  

For convergence analysis to the exact invariant optimizer we consider two cases: (i) sector-based and (ii) non-sector-based nonlinearities. Consider the residual function $\overline{F} = F(\mb{x}(k))-F(\mb{x}^*)$. Following the  \cite{mrd_2020fast,mrd20211st}, it can be shown that this residual function is \textit{non-increasing} for \textit{sign-preserving} nonlinearities (i.e., nonlinear mappings satisfying $\mb{z}h(\mb{z}) \geq 0$) and  \textit{decreasing} for \textit{strongly sign-preserving} nonlinearities satisfying $\mb{z}h(\mb{z}) > 0$ for $\mb{z} \neq 0$. For the latter, the convergence to the exact optimizer can be proved using discontinuous Lyapunov analysis and the convergence rate can be defined under certain conditions. For example, assuming strongly convex and smooth costs (with Lipschitz gradients) satisfying $v\leq \partial_x^2 f_i(\mb{x}_i) \leq u$ and any nonlinearity satisfying $\underline{\alpha} \leq \frac{h(\mb{z})}{\mb{z}} \leq \overline{\alpha}$
we have,
\begin{align} \label{eq_cov_rate}
    \frac{\overline{F}(k+1)}{\overline{F}(k)} \leq 1- \overline{\eta}v(\underline{\alpha}\lambda_2 - \frac{u}{2}\lambda_n^2\overline{\alpha} \overline{\eta})
\end{align} 
with $\lambda_2$ and $\lambda_n$ as the second smallest and largest eigenvalue of the connected MAN. Recall that, for connected networks, $\lambda_2$ represents the algebraic connectivity (Fiedler value) as a measure of link density of the network, i.e., for highly connected networks (and higher $\lambda_2$ value) the convergence to the optimizer is faster (according to \eqref{eq_cov_rate}). Note that the above results on convergence hold for certain bound on the step-rate $\overline{\eta}$, i.e., for very large values of $\overline{\eta}$ the optimization algorithm may diverge. The convergence rate can be extended to uniformly connected networks over period $T$, i.e., one can define the rate for $\frac{\overline{F}(k+T)}{\overline{F}(k)}$.  

In nonlinear systems theory, the condition  $\underline{\alpha} \leq \frac{h(\mb{z})}{\mb{z}} \leq \overline{\alpha}$ refers to sector-based nonlinearities. In quantized systems example, \textit{logarithmic} quantization represents such a nonlinearity, while \textit{uniform} quantization
is non-sector-based. It can be shown that all sector-based nonlinearities with $\underline{\alpha}>0$ are ``strongly'' sign-preserving and, thus, the distributed solutions \eqref{eq_dt_nonlin1}-\eqref{eq_dt_nonlin2} exactly converge to the optimizer $\mb{x}^*$, while for the non-sector-based case this may not necessarily hold. For example, for uniform quantization one can prove convergence only to an $\varepsilon$-neighborhood of the optimizer $\|\mb{x}(k)-\mb{x}^*\|\leq \varepsilon$, where $\varepsilon$ is a function of the quantization level $\delta$. This implies that the solution always has some non-zero residuals in the steady-state. For some other example nonlinearities, e.g., non-Lipschitz sign-based solutions, where $\overline{\alpha} \rightarrow \infty$ the exact optimizer cannot be reached (in DT) due to the so-called \textit{chattering} phenomena which results in unwanted oscillations around the optimizer (which can be reduced by choosing smaller steps). However, such issues can be avoided by using the CT counterpart of the dynamics, for example, to reach fixed-time convergence in the economic dispatch problem, see the next section for the illustrative simulation. This is also addressed in machine learning applications and distributed support-vector-machines for classification over MAN \cite{csl2021}.

\section{Some Example Applications and Simulations}
This section provides some simplified models as academic examples illustrating how the mentioned nonlinearities can be addressed in some existing finite-sum optimization problems.   

\subsection{CPU Scheduling in DT with Quantized Communications}
\begin{figure} [t]
		\centering
        \raisebox{-0.5\height}{\includegraphics[width=1.5in]{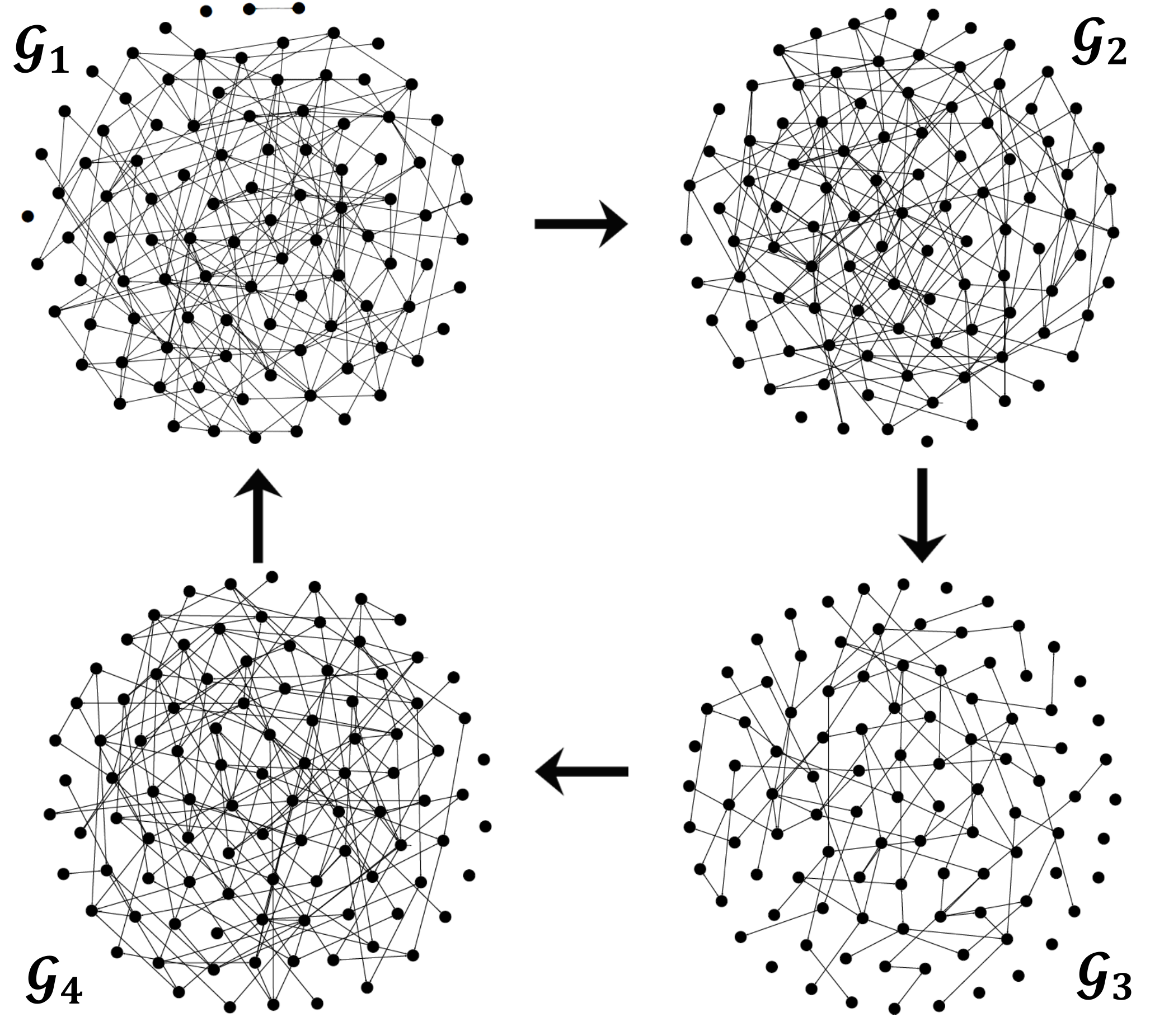}} 
        \raisebox{-0.5\height}{\includegraphics[width=0.9in]{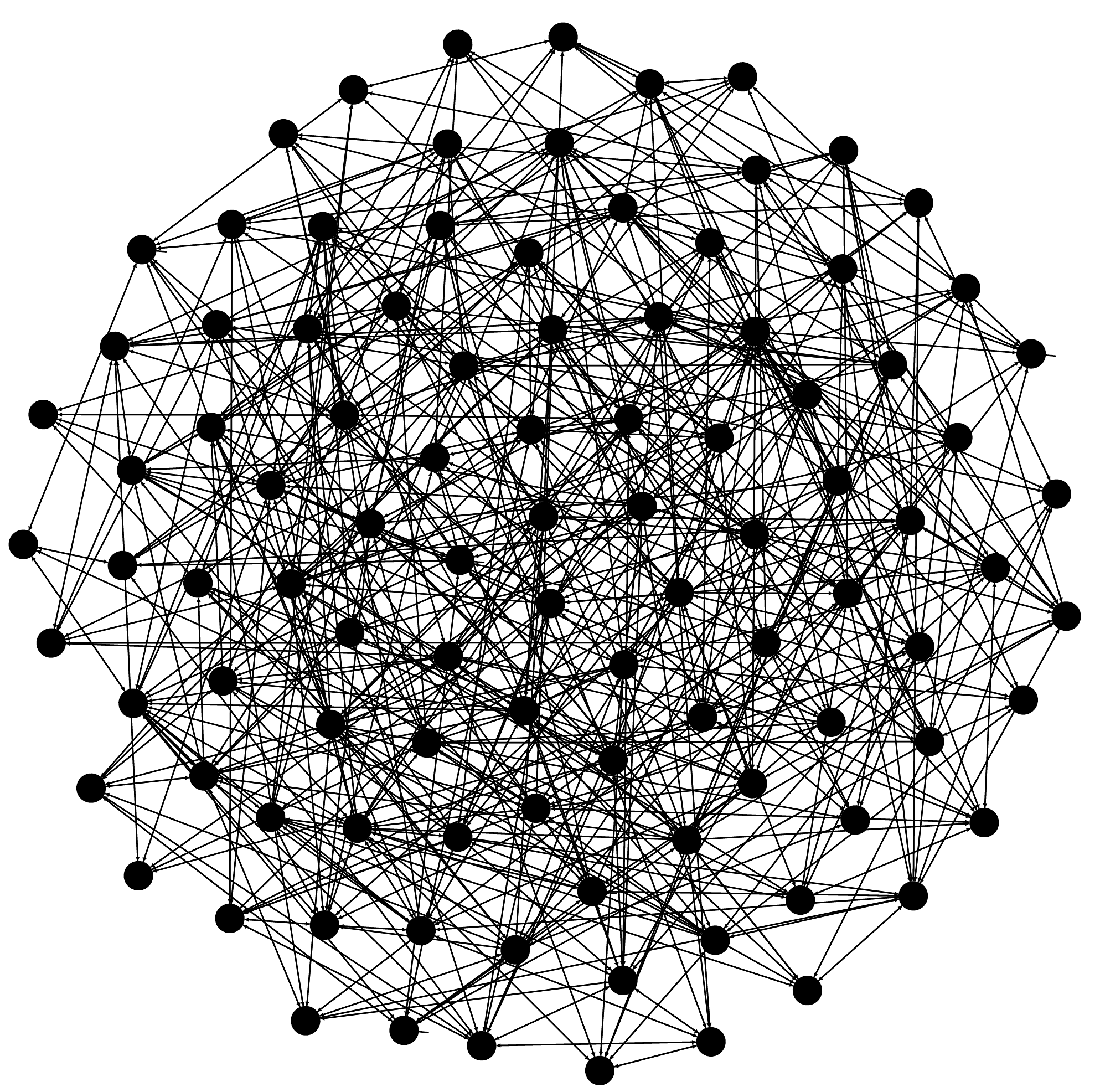} }
		\caption{This figure shows the dynamic network of $n=100$ CPU nodes for the simulation. Although, the networks (on the left) are disconnected, their union (on the right) is connected. This refers to the notion of uniform-connectivity, which is the least connectivity requirement for convergence. } 
		\label{fig_graph} 
	\end{figure}

\begin{figure*} [t]
		\centering
		\includegraphics[width=2.3in]{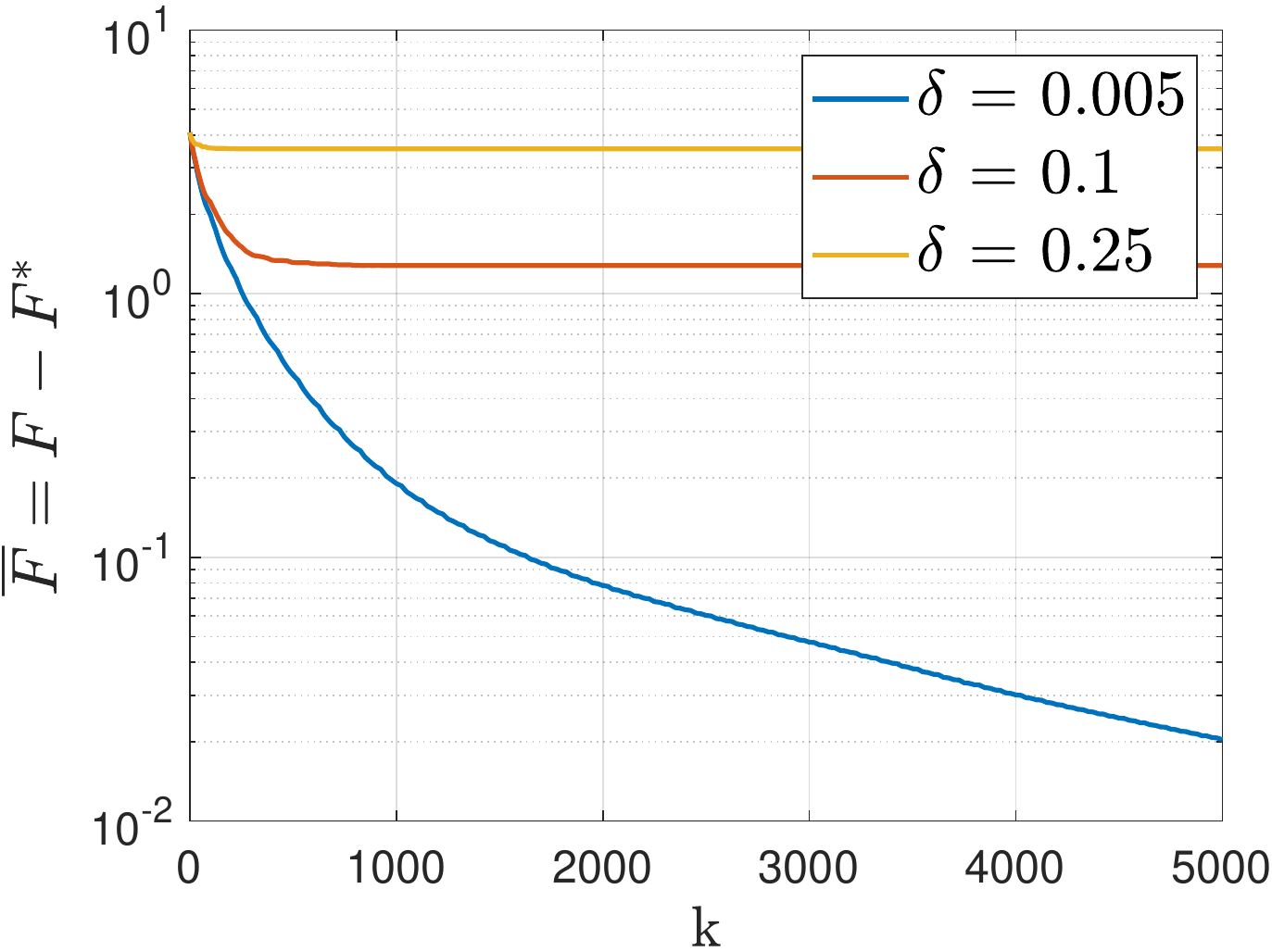}
 		\includegraphics[width=2.3in]{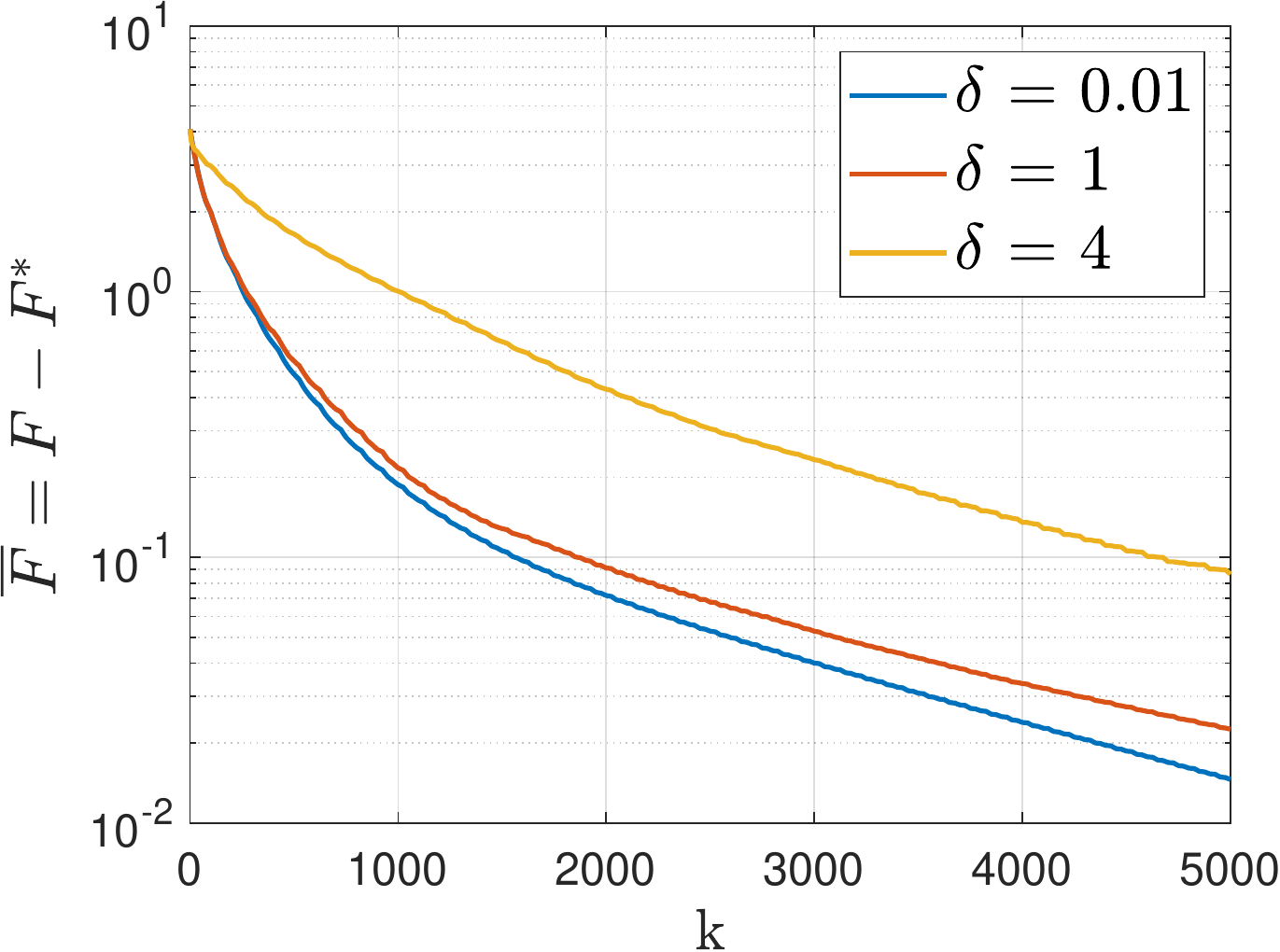}
 		\includegraphics[width=2.22in]{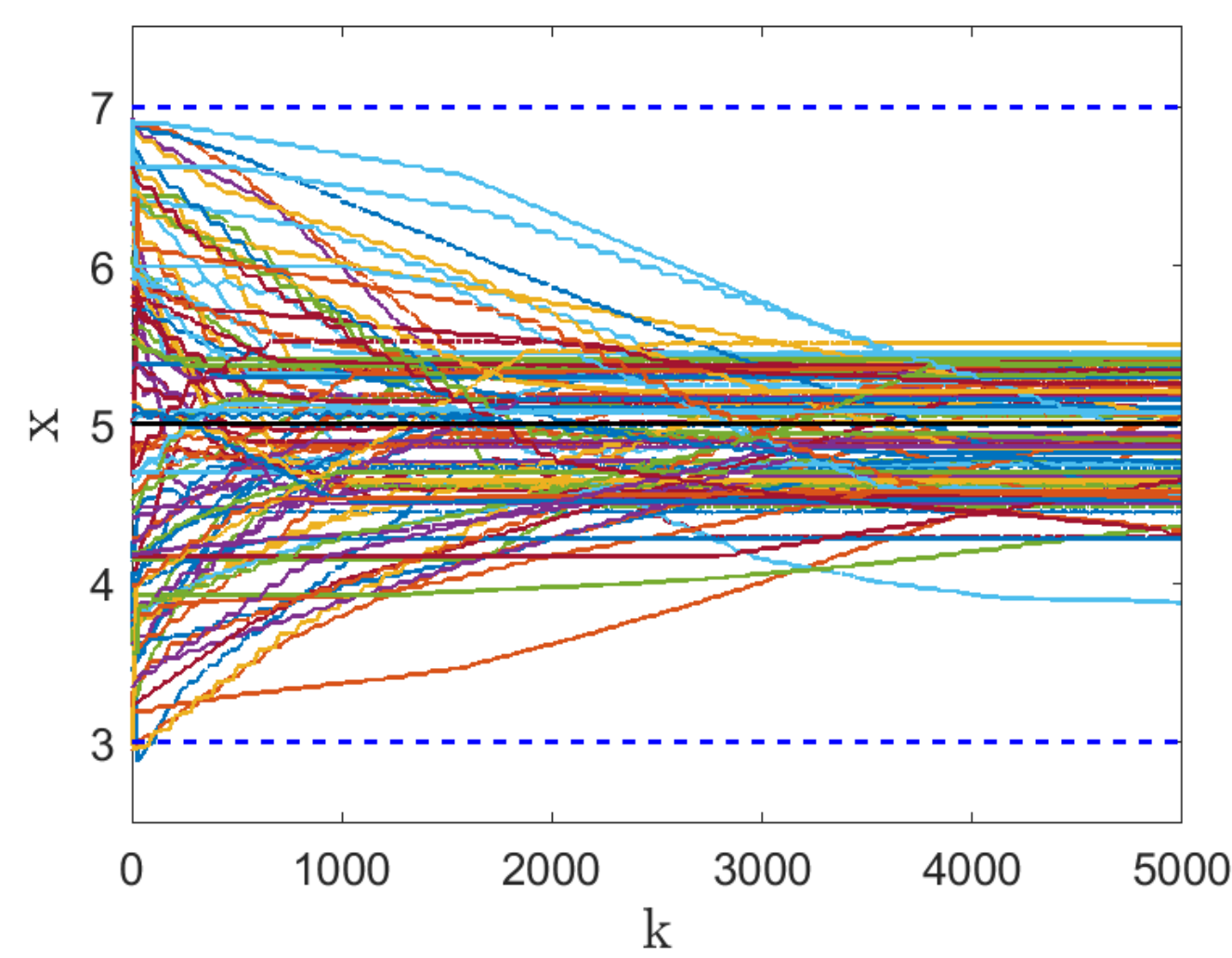}
		\caption{This figure shows the residuals under uniform quantizer (left) and logarithmic quantizer (right) for different quantization levels $\delta$. Logarithmic quantizer (as a sector-based nonlinearity)  is ``strongly'' sign-preserving, while uniform quantizer is only sign-preserving (but not strongly) which results in steady-state residuals depending on the value of $\delta$. The right figure shows the time-evolution of the workloads towards optimal state $\mb{x}^*$ under logarithmic-quantizer protocol ($\delta=4$).
		}
		\label{fig_quant}
\end{figure*}

Optimizing the workloads over network of CPUs (or computing servers) is considered as an example here.
The local cost of each CPU is defined in quadratic form as \cite{rikos2021optimal,kalyvianaki2009self,charalambous2010min}, 
\begin{align}\label{eq:fiz}
f_i(\mb{x}_i) = \frac{1}{2}\pi_i \left(\mb{x}_i- \frac{\rho_{i}+u_{i}}{\pi_i} \right)^2
\end{align}
with random parameters $\pi_i \in (0~0.1], \frac{\rho_{i}+u_{i}}{\pi_i} \in [4.5~5.5]$ associated with CPU at node $i$. For the simulation, we consider the sum of the workloads as $K=1000$; max and min workloads at every computing server are $7$ and $3$ (the box constraints). The chosen numerical values are only for the sake of simulation and may not necessarily follow the specifications in \cite{rikos2021optimal,kalyvianaki2009self,charalambous2010min}.
As an academic example, a random dynamic network of $n=100$ nodes switching between the $4$ networks in Fig.~\ref{fig_graph} is used for this simulation. Quantized data exchanges (both logarithmic and uniform) are considered over the network channels with different quantization levels (for different possible bit rates over the communication channels), and the results are shown and compared in Fig.~\ref{fig_quant} in terms of convergence rate and accuracy (convergence to the exact optimizer). 

\subsection{Fast Coordination in CT Nonlinear Systems: Application to Smart Grid}
\begin{figure*} [t]
		\centering
        \includegraphics[width=1.75in]{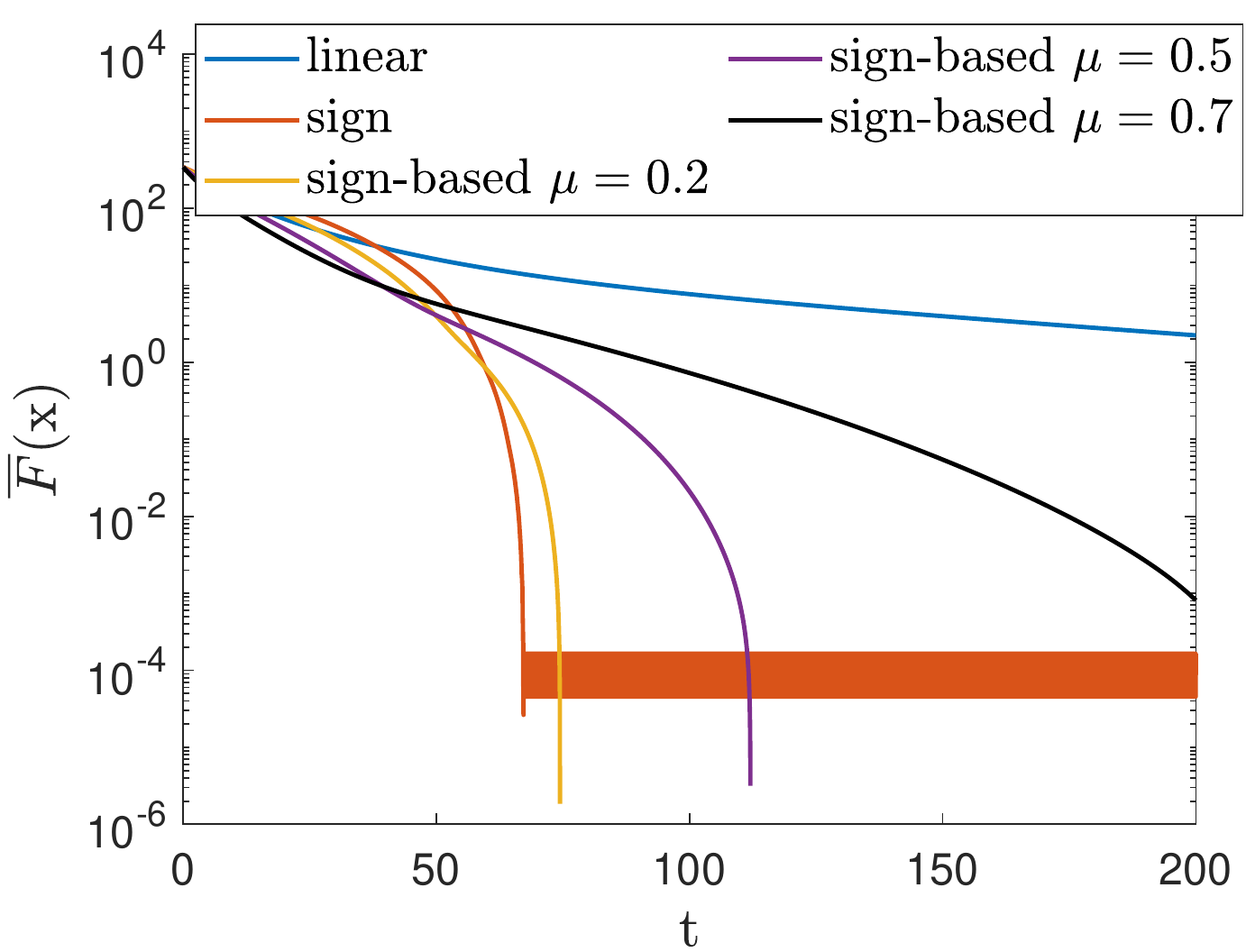}
        \includegraphics[width=1.75in]{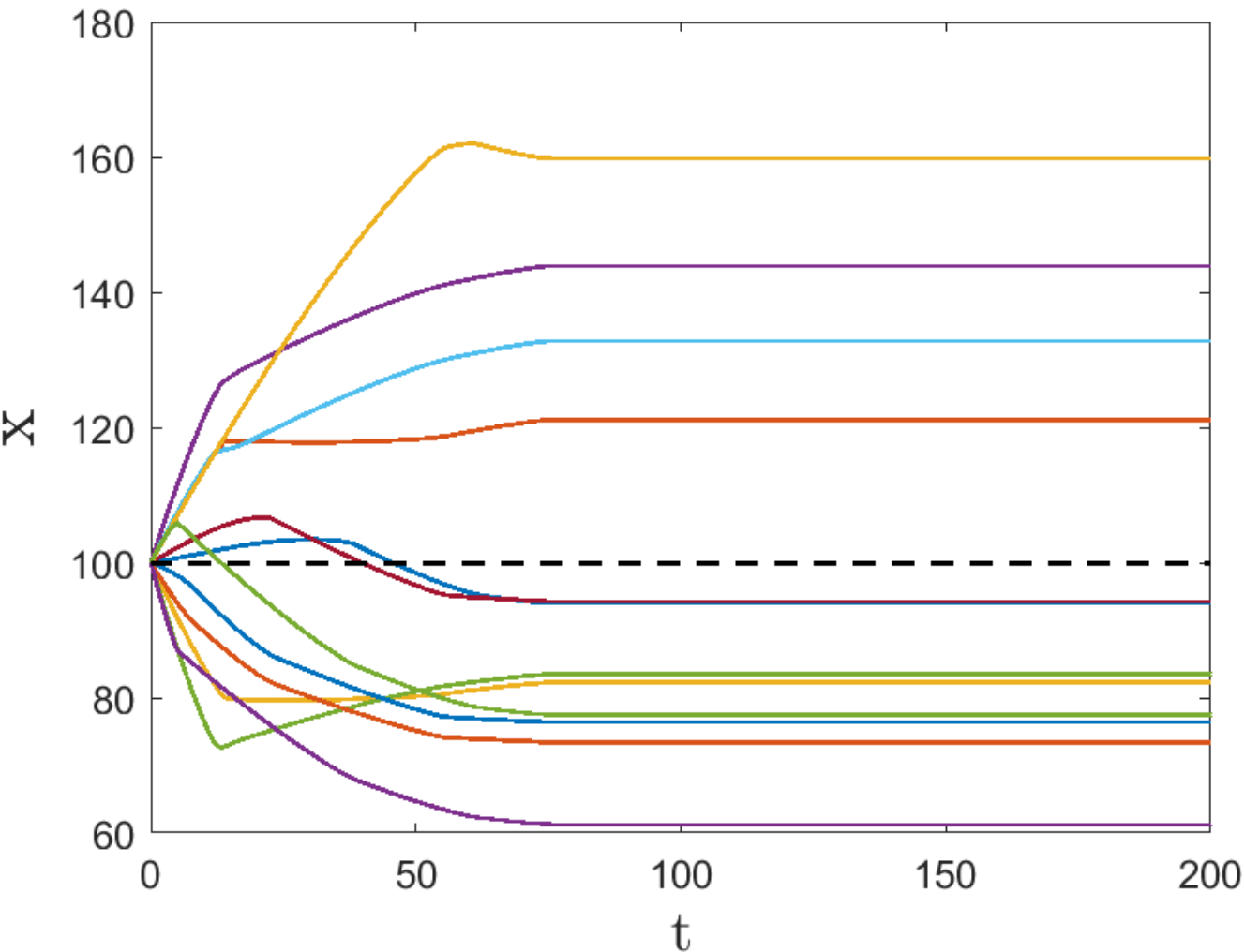}
        \includegraphics[width=1.75in]{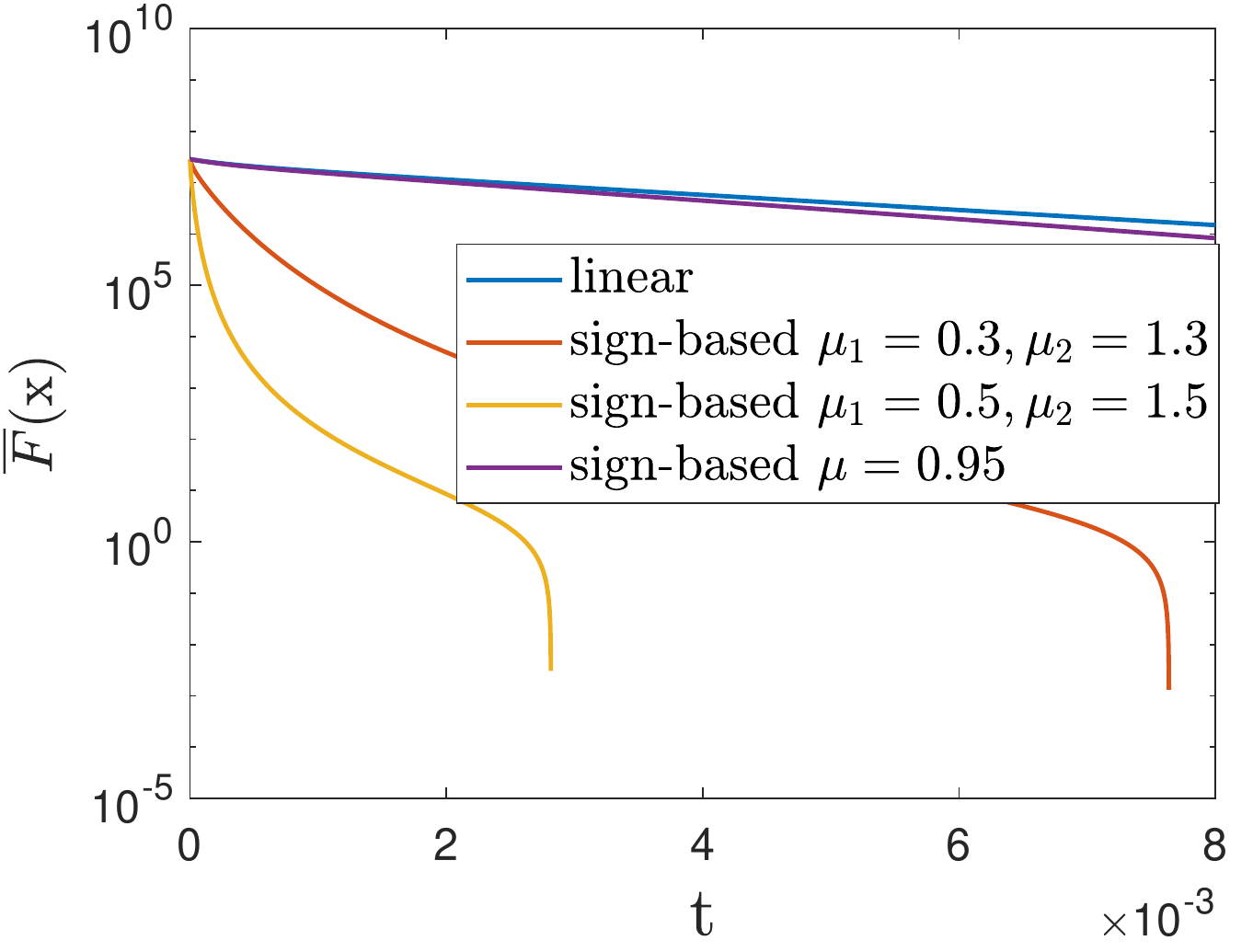}
        \includegraphics[width=1.75in]{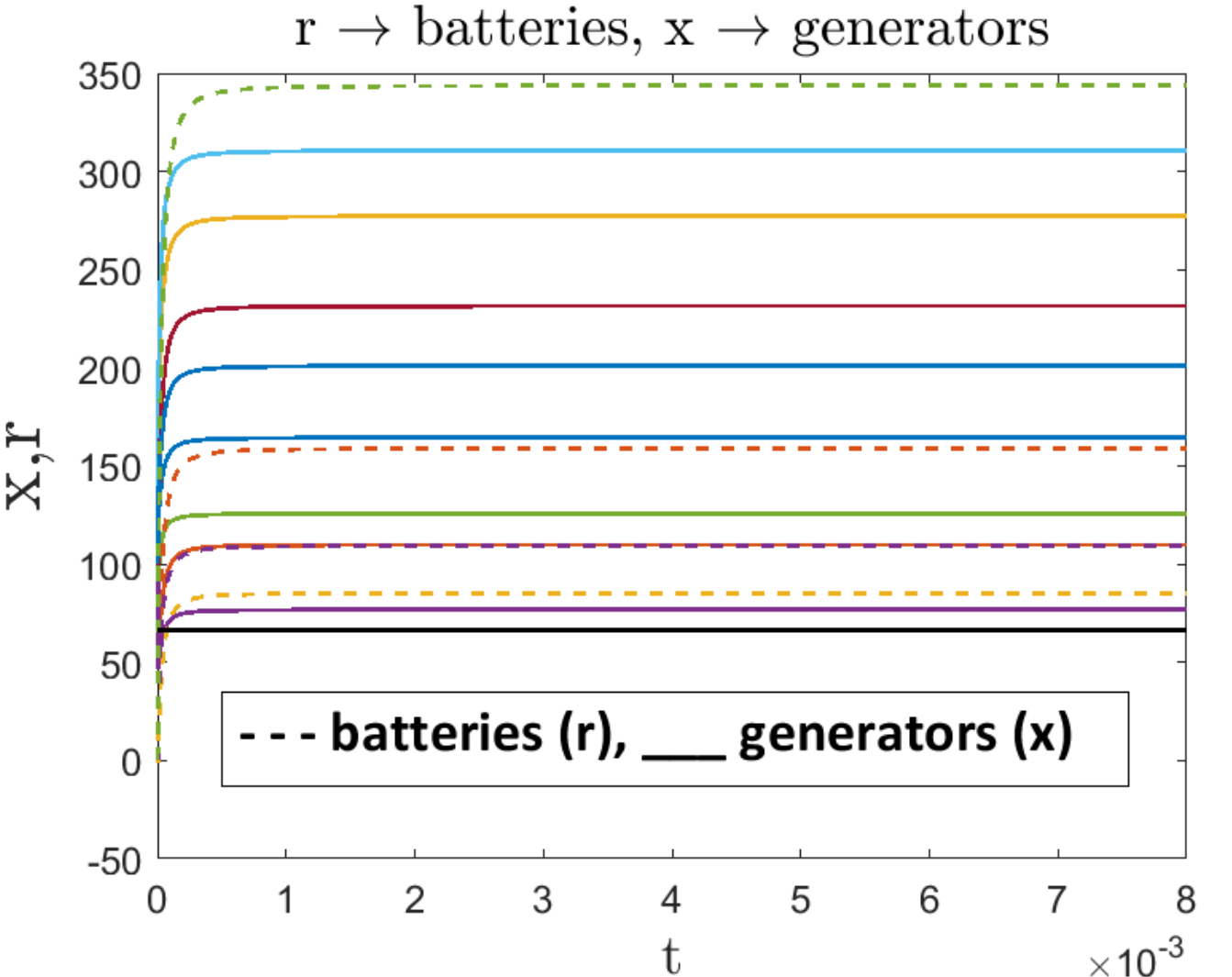}
		\caption{This figure shows how to improve the convergence rate. (Left) Adding sign-based nonlinearities improves the rate of the linear solution for power allocation over a network of generators (blue). However, it may add unwanted chattering around the optimizer $\mb{x}^*$ and optimal cost $F^*$ for discontinuous nonlinearities. One can instead use smooth and continuous non-Lipschitz dynamics. (Second-left) Convergence of the power states (to be assigned to the generators) over finite time (in this case before $t=100$ for $\mu=0.2$) while preserving the average (and the sum) constant (black dashed line). (Second-right) This figure further improves the convergence rate of the power allocation (for cost function \eqref{eq_f_quad2}) by using complementary sign-based nonlinearities with parameters $0\leq \mu_1 < 1$ and $ \mu_2 > 1$. (Right) The evolution of power states for generators (solid) and batteries (dashed) are shown (for $\mu_1=0.5 $ and $ \mu_2=1.5$) with the black solid line as  $\frac{D}{n}$ (with $D$ as the difference of total power generation and total power reservation).}
		\label{fig_edp}
\end{figure*}

The application in networks of power generators for efficient power production over the smart grid is considered to improve the linear results in \cite{cherukuri2015tcns}. 
The power generation cost at all the generators typically follows the form, 
\begin{eqnarray} \label{eq_f_quad}
\sum_{i=1}^n f_i(\mb{x}_i) =  \sum_{i=1}^n \gamma_i \mb{x}_i^2+ \beta_i \mb{x}_i + \alpha_i,~\mbox{s.t.}~\sum_{i=1}^n \mb{x}_i = D
\end{eqnarray}
with the parameters of the cost functions defined based on the type of the power generators (coal-fired, oil-fired, etc.), for example, as given in Table~\ref{tab_par} \cite{chen2018fixed}.
\begin{table} [h]
	\centering
	\caption{Parameters of the generator cost functions.}
	\begin{tabular}{|c|c|c|c|}
		\hline
		& $\alpha_i$ (\$/$h$) & $\beta_i$ (\$/$MWh$) & $\gamma_i$ (\$/$MW^2h$)\\
		\hline
		Type-A & 561& 2.0 & 0.04\\
		\hline
		Type-B& 310& 3.0& 0.03\\
		\hline
		Type-C &78 & 4.0& 0.035 \\
		\hline
		Type-D& 561& 4.0& 0.03\\
		\hline
		Type-E &78 & 2.5& 0.04 \\
		\hline		
	\end{tabular}
	\label{tab_par}
\end{table}
Adding nonlinear non-Lipschitz mappings to the linear Laplacian-dynamics in \cite{cherukuri2015tcns} significantly improves the convergence rate. We consider the CT nonlinear actuation dynamics  \eqref{eq_ct_nonlin1} in the form,
\begin{align} \label{eq_ct_sgn}
    \dot{\mb{x}}_i &=  \sum_{j\in \mc{N}_i}W_{ji} \mbox{sgn}^\mu(\partial_x f_j(t)-\partial_x f_i(t))
\end{align}
with $\mbox{sgn}^\mu(\mb{z})=\mb{z}|\mb{z}|^{\mu-1}$. Different values for $\mu$ are considered, $\mu=0$ (sign function), $\mu=0.2$, $\mu=0.5$, $\mu=0.7$. The results over a simple cyclic network of size $n=12$ and demand load of $D=1200$ are shown in Fig.~\ref{fig_edp}. From the figure, one can see that the sign function reaches faster convergence (however, with chattering at steady-state due to non-Lipschitz property). As a remedy, one can consider small value of $\mu=0.2$ (instead of $\mu=0$) to improve the chattering phenomena while keeping the convergence fast enough.
The results can be extended to add load-based reserve capacities \cite{vrakopoulou2017chance}. This setup further provides chance-constrained optimal power ﬂow to schedule both generation and
reserves from some wind generators and aggregations of controllable electric loads. The controllable loads are modelled as \textit{thermal energy storage units} with a redispatch mechanism to manage the energy state (as a battery’s state of charge), i.e., a battery type of dynamics is considered with no efficiency issues as it refers to a virtual battery that is based on thermostatically controlled loads, see details in \cite{vrakopoulou2017chance}. The revised mathematical formulation is\footnote{Some additional deterministic/probabilistic constraints are further given in \cite{vrakopoulou2017chance}. In this paper, we only consider a simplified version to address the effect of model nonlinearities on the convergence over a \textit{distributed} setup.},   
\begin{align} \label{eq_f_quad2}
\sum_{i=1}^n f_i(\mb{x}_i) &=  \sum_{i=1}^n \gamma_i \mb{x}_i^2+ \beta_i \mb{x}_i + \alpha_i + \sum_{j=1}^{n_r} c_j\mb{r}_j \\ &~\mbox{s.t.}~\sum_{i=1}^n \mb{x}_i = D +\sum_{j=1}^{n_r} \mb{r}_j  
\end{align}
where the additive terms $\mb{r}_i$ on the load demand-side represent the states of $n_r$ charging batteries with cost-related coefficients $c_j$. In this setup the additive power production by the generators is reserved by the batteries.  
The cost is strongly convex and solvable via distributed algorithms in Section~\ref{sec_solution} by introducing new set of $n+n_r$ variables $\mb{y}=[\mb{x};-\mb{r}]$. Following the KKT conditions, the optimal state $\mb{y}^*$ is such that $\nabla F(\mb{y}^*) \in \mbox{span}\{\mb{1}_{n+n_r}\}$ (or equivalently $\nabla F([\mb{x}^*;\mb{r}^*]) \in \mbox{span}\{[\mb{1}_{n};-\mb{1}_{n_r}\}$). Repeating the same simulation (with 
different cost parameters over a 2-hop network) for $n_r=4$ thermal batteries (reserving) and $n=8$ generators (producing power), the optimal power allocation is shown in Fig.~\ref{fig_edp} for $D=800$. To reach even faster convergence, two \textit{complementary} sign-based terms with $0<\mu_1<1$ and $1<\mu_2$ are considered (termed as fixed-time protocols \cite{mrd_2020fast}) and the convergence rate is compared with the linear model.   

\section{Conclusions}
We analyze \textit{distributed} techniques for optimization over nonlinear channels and nonlinear actuators with possible application to finite-sum resource allocation over MAN. We discuss the conditions ensuring convergence and feasibility of the coupled solutions among the agents; and, in particular, we show by simulation that (i) the exact optimality is ensured for sector-based (and in general ``strongly'' sign-preserving) nonlinearities, while (ii) for non-sector-based (and sign-preserving) nonlinear models one can \textit{only} guarantee convergence to the $\varepsilon$-neighborhood of the optimizer (i.e., certain steady-state residuals). 

\bibliographystyle{IEEEbib}
\bibliography{bibliography}
\end{document}